\documentclass{PoS}

\usepackage{amsmath,amssymb}
\usepackage{bbm}
\usepackage{dcolumn}

\newcommand{\1}{\mathbbm{1}}
\newcommand{\eps}{\varepsilon}

\DeclareMathOperator{\diag}{diag}
\DeclareMathOperator{\re}{Re}
\DeclareMathOperator{\sign}{sign}
\DeclareMathOperator{\tr}{tr}
\DeclareMathOperator{\ind}{index}

\title{Nonzero chemical potential in the overlap Dirac operator and comparison to random matrix theory}

\ShortTitle{Nonzero chemical potential in the overlap Dirac operator and random matrix theory}

\author{\speaker{Jacques Bloch} and Tilo Wettig\\
        Institute for Theoretical Physics, University of
  Regensburg, 93040 Regensburg, Germany\\
        E-mail: \email{jacques.bloch@physik.uni-regensburg.de}}

\abstract{
  In this talk we present the results published recently in
Ref.\ \cite{Bloch:2006cd}, where we showed how to introduce a quark 
chemical potential in the overlap Dirac operator.
The resulting operator satisfies a Ginsparg-Wilson
  relation and has exact zero modes.  It is no longer
  $\gamma_5$-Hermitian, but its nonreal eigenvalues still occur in
  pairs.  We compute the spectral density of the operator on the
  lattice and show that, for small eigenvalues, the data agree with
  analytical predictions of non-Hermitian chiral random matrix theory
  for both trivial and nontrivial topology.
  We also explain an observed change in the number of zero modes as a
  function of chemical potential.
}

\FullConference{XXIVth International Symposium on Lattice Field Theory\\
                July 23-28, 2006\\
                Tucson, Arizona, USA}

\begin{document}

Quantum chromodynamics (QCD) at nonzero baryon density is important for the study of relativistic heavy-ion collisions, neutron stars, and the early universe \cite{owe}. 
The effects of such a density are investigated by introducing a quark chemical potential $\mu$ in the QCD Dirac operator.

For $\mu\ne0$ the Dirac operator loses its hermiticity properties and its
spectrum moves into the complex plane.  This causes a variety of
problems, both analytically and numerically.  Lattice simulations are
the main source of nonperturbative information about QCD, but at
$\mu\ne0$ they cannot be performed by standard importance sampling
methods because the measure of the functional integral, which
includes the complex fermion determinant, is no longer positive
definite.  

A better analytical understanding of QCD at very high baryon density
has been obtained by a number of methods \cite{Al02}, and the QCD
phase diagram has been studied in model calculations based on
symmetries \cite{HJSSV98}.  Chiral random matrix theory (RMT)
\cite{SV93}, which makes exact analytical predictions for the
correlations of the small Dirac eigenvalues, has been extended to
$\mu\ne0$ \cite{Step96}, and a mechanism was identified \cite{OSV05}
by which the chiral condensate at $\mu\ne0$ is built up from the
spectral density of the Dirac operator in the
complex plane, in stark contrast to the Banks-Casher mechanism at
$\mu=0$. This mechanism and its relation to the sign problem was also discussed by Splittorff in a plenary talk at this conference \cite{KimLat06}.

A first comparison of lattice data with RMT predictions at $\mu\ne0$
was made in Ref.~\cite{AW04} using staggered fermions.  One issue with
staggered fermions is that the topology of the gauge field is only
visible in the Dirac spectrum if the lattice spacing is small and
various improvement and/or smearing schemes are applied
\cite{stag-top}.  To avoid these issues, we would like to work with a
Dirac operator that implements a lattice version of chiral symmetry
and has exact zero modes at finite lattice spacing \cite{Ha04}.  The overlap
operator \cite{overlap} satisfies these requirements at $\mu=0$.  In
the following, we show how the overlap operator can be modified to
include a nonzero quark chemical potential\footnote{See also Ref.\ \cite{BW98} for a perfect lattice action at $\mu\ne0$ and Ref.\ \cite{BH00} for an overlap-type operator at $\mu\ne0$ in momentum space.}
\cite{Bloch:2006cd}.  We then
study the spectral properties of this operator as a function of $\mu$
and compare data from lattice simulations with RMT predictions.  As we
shall see, the overlap operator has exact zero modes also at nonzero
$\mu$, which allows us, for the first time, to test predictions of
non-Hermitian RMT for nontrivial topology.

We begin with the well-known definition of the Wilson Dirac operator
$D_W$ including a chemical potential $\mu$ \cite{hk83},
\begin{align}
  D_W(\mu)&=\1-\kappa\sum_{i=1}^3\left(T_i^++T_i^-\right)
  -\kappa\left(e^\mu T_4^++e^{-\mu} T_4^-\right)\:,\notag\\
  (T_\nu^\pm)_{yx}&=(1\pm\gamma_\nu)U_{\pm\nu}(x)\delta_{y,x\pm\hat\nu}\:,
  \label{eq:dwmu}
\end{align}
where $\kappa=1/(2m_W+8)$ with the Wilson mass $m_W$, the
$U\in\text{SU(3)}$ are the lattice gauge fields, and the $\gamma_\nu$
are the usual Euclidean Dirac matrices.  Unless displayed explicitly,
the lattice spacing $a$ is set to unity.  

The overlap operator is defined at $\mu=0$ by \cite{overlap}
\begin{align}
  \label{eq:overlap0}
  D_\text{ov}(0)=\1+\gamma_5\eps(\gamma_5D_W(0))\:,
\end{align}
where $\eps$ is the matrix sign function and
$\gamma_5=\gamma_1\gamma_2\gamma_3\gamma_4$.  $m_W$ must be in the
range $(-2,0)$ for $D_\text{ov}(0)$ to describe a single Dirac fermion
in the continuum.  The properties of $D_\text{ov}(0)$ have been
studied in great detail in the past years.  In particular, its
eigenvalues are on a circle in the complex plane with center at
$(1,0)$ and radius 1, its nonreal eigenvalues come in complex
conjugate pairs, and it can have exact zero modes without fine-tuning.
$D_\text{ov}(0)$ satisfies a Ginsparg-Wilson relation \cite{GW82} of
the form
\begin{align}
  \label{eq:gw}
  \{D,\gamma_5\}=D\gamma_5 D\:.
\end{align}

We now extend the definition of the overlap operator to $\mu\ne0$.
The operator $D_W(0)$ in Eq.~\eqref{eq:overlap0} is
$\gamma_5$-Hermitian, i.e., $\gamma_5D_W(0)\gamma_5=D_W^\dagger(0)$,
and therefore the operator $\gamma_5D_W(0)$ in the matrix sign
function is Hermitian.  However, for $\mu\ne0$, $D_W(\mu)$ is no
longer $\gamma_5$-Hermitian.  Defining the overlap operator at nonzero
$\mu$ by
\begin{align}
  \label{eq:overlapmu}
  D_\text{ov}(\mu)=\1+\gamma_5\eps(\gamma_5D_W(\mu))\:,
\end{align}
we now need the sign function of a non-Hermitian matrix.  In general, a
function $f$ of a non-Hermitian matrix $A$ can be defined by the contour
integral
\begin{align}
f(A) = \frac{1}{2\pi i} \oint_\Gamma dz \; f(z) (z \1-A)^{-1} \,,
\label{eq:ointfA}
\end{align}
where the spectrum of $A$ is enclosed by the contour $\Gamma$ and the matrix integral is defined on an element-by-element basis.
A more convenient expression can be obtained if $A$ is
diagonalizable.  In this case we can write $A=U\Lambda U^{-1}$, where
$U\in\text{Gl}(N,\mathbb{C})$ with $N=\dim(A)$ and
$\Lambda=\diag(\lambda_1,\ldots,\lambda_N)$ with
$\lambda_i\in\mathbb{C}$.  Then, applying Cauchy's theorem to Eq.~\eqref{eq:ointfA}, $f(A)=Uf(\Lambda)U^{-1}$, where
$f(\Lambda)$ is a diagonal matrix with elements $f(\lambda_i)$. In particular,  the matrix sign function can be defined by \cite{sign}
\begin{align}
  \label{eq:sign}
  \eps(A)=U\sign(\re\Lambda)U^{-1}\:.
\end{align}
This definition ensures that $\eps^2(A)=\1$ and gives the correct
result if $\Lambda$ is real.  An equivalent definition is
$\eps(A)=A(A^2)^{-1/2}$ \cite{H94}.  Eqs.~\eqref{eq:overlapmu} and
\eqref{eq:sign} constitute our definition of $D_\text{ov}(\mu)$.  The
sign function is ill-defined if one of the $\lambda_i$ lies on the
imaginary axis.  Also, it could happen that $\gamma_5D_W(\mu)$ is not
diagonalizable (one would then resort to a Jordan block
decomposition).  Both of these cases are only realized if one or more
parameters are fine-tuned, and are unlikely to occur in realistic
lattice simulations.

It is relatively straightforward to derive the following properties of
$D_\text{ov}(\mu)$:\vspace{-3mm}
\begin{itemize}\itemsep-1.0mm
\item It is no longer $\gamma_5$-Hermitian.

\item It still satisfies the Ginsparg-Wilson relation
  \eqref{eq:gw} because of $\eps^2(A)=\1$.  Thus, we still have a
  lattice version of chiral symmetry, and the operator has exact zero
  modes without fine-tuning.
  
\item Its eigenvalues not equal to 0
  or 2 no longer come in complex conjugate pairs, but every such
  eigenvalue $\lambda$ (with eigenvector $\psi$) comes with a
  partner $\lambda/(\lambda-1)$ (with eigenvector $\gamma_5\psi$).

\item Its eigenvectors corresponding to
  eigenvalues 0 or 2 can be arranged to have definite chirality. 

\end{itemize}

We now turn to our (quenched) lattice simulations.  The computation of
the sign function of a non-Hermitian matrix is very demanding.  We are
currently investigating various approximation schemes, but
in this initial study we decided to compute the sign function and to
diagonalize $D_\text{ov}(\mu)$ exactly using LAPACK.  For the
comparison with RMT we need high statistics, which restricts us to a
very small lattice size.  We have chosen the same parameter set as in
Ref.~\cite{ehkn99} to be able to compare with previous results at
$\mu=0$.  The lattice size is $V=4^4$, the coupling in the standard
Wilson action is $\beta=5.1$, the Wilson mass is $m_W=-2$, and the
quark mass is $m_q=0$.  

\begin{figure}[-t]
  \centerline{\includegraphics[height=50mm]{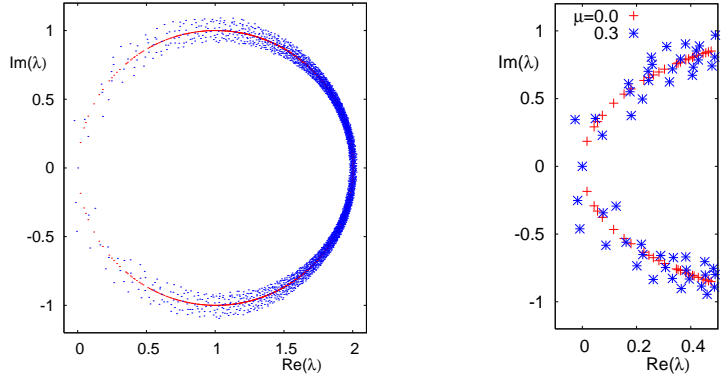}}
  \caption{Spectrum of $D_\text{ov}(\mu)$ for $\mu=0$ and $\mu=0.3$
    for a typical configuration.  The figure on the right is a
    magnification of the region near zero.}
  \label{fig:spectrum}
\end{figure}

\begin{figure}[b]
\centerline{
\parbox{8cm}{\includegraphics[height=50mm]{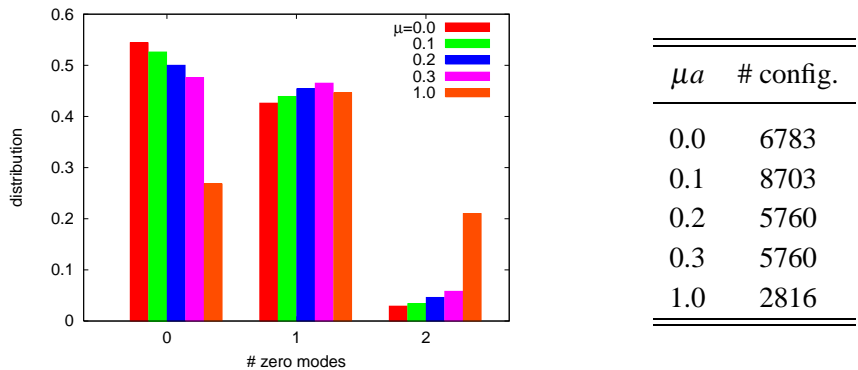}}
\hspace{5mm}
 \parbox[b]{3cm}{
     \begin{tabular}{cc}
	 \hline\hline\\[-4mm]
       $\mu a$ & \# config.
       \\[1mm]\hline\\[-3mm]
       0.0 & 6783\\
       0.1 & 8703\\
       0.2 & 5760\\
       0.3 & 5760\\
       1.0 & 2816\\
	   \hline\hline
     \end{tabular}}
  	 }
\caption{Distribution of the number of zero mores of $D_\text{ov}(\mu)$ for $\mu=0, 0.1, 0.2, 0.3, 1.0$, and number of configurations (right table).}
\label{table1}
\end{figure}

In Fig.~\ref{fig:spectrum} we show the spectrum of $D_\text{ov}(\mu)$ for
a typical configuration for $\mu=0$ and $\mu=0.3$.  As expected, we
see that the eigenvalues move away from the circle as $\mu$ is turned
on.  Another observation is that the number of zero modes of
$D_\text{ov}(\mu)$ for a given configuration can change as a function
of $\mu$, see Figure~\ref{table1}.  This can be understood from the
relation between the anomaly and the index of $D_\text{ov}$
\cite{anomaly1,anomaly2},
\begin{align}
  \label{eq:index}
  -\tr(\gamma_5D_\text{ov})=2\ind(D_\text{ov})\:,
\end{align}
which we can show to remain valid at $\mu\ne0$.  Using
$\tr(\gamma_5D_\text{ov})=\tr[\eps(\gamma_5D_W)]$ and the fact that the eigenvalues of the sign
function are $+1$ or $-1$, one has $\ind(D_\text{ov})=(n_-^W-n_+^W)/2$,
where $n_\pm^W$ denotes the number of eigenvalues of
$\gamma_5D_W(\mu)$ with real part $\gtrless0$. Therefore the number of zero modes for a given configuration is determined by the difference of the number of eigenvalues of $\gamma_5 D_W$ lying left and right of the imaginary axis.
As $\mu$ changes, an
eigenvalue of $\gamma_5D_W$ can move across the imaginary axis.  As a
result, 
$\ind(D_\text{ov})$ changes by 1, which explains the observation.  We believe that this is a lattice artefact which will disappear in the continuum limit.

\begin{figure}
\centerline{
\includegraphics[width=5.5cm]{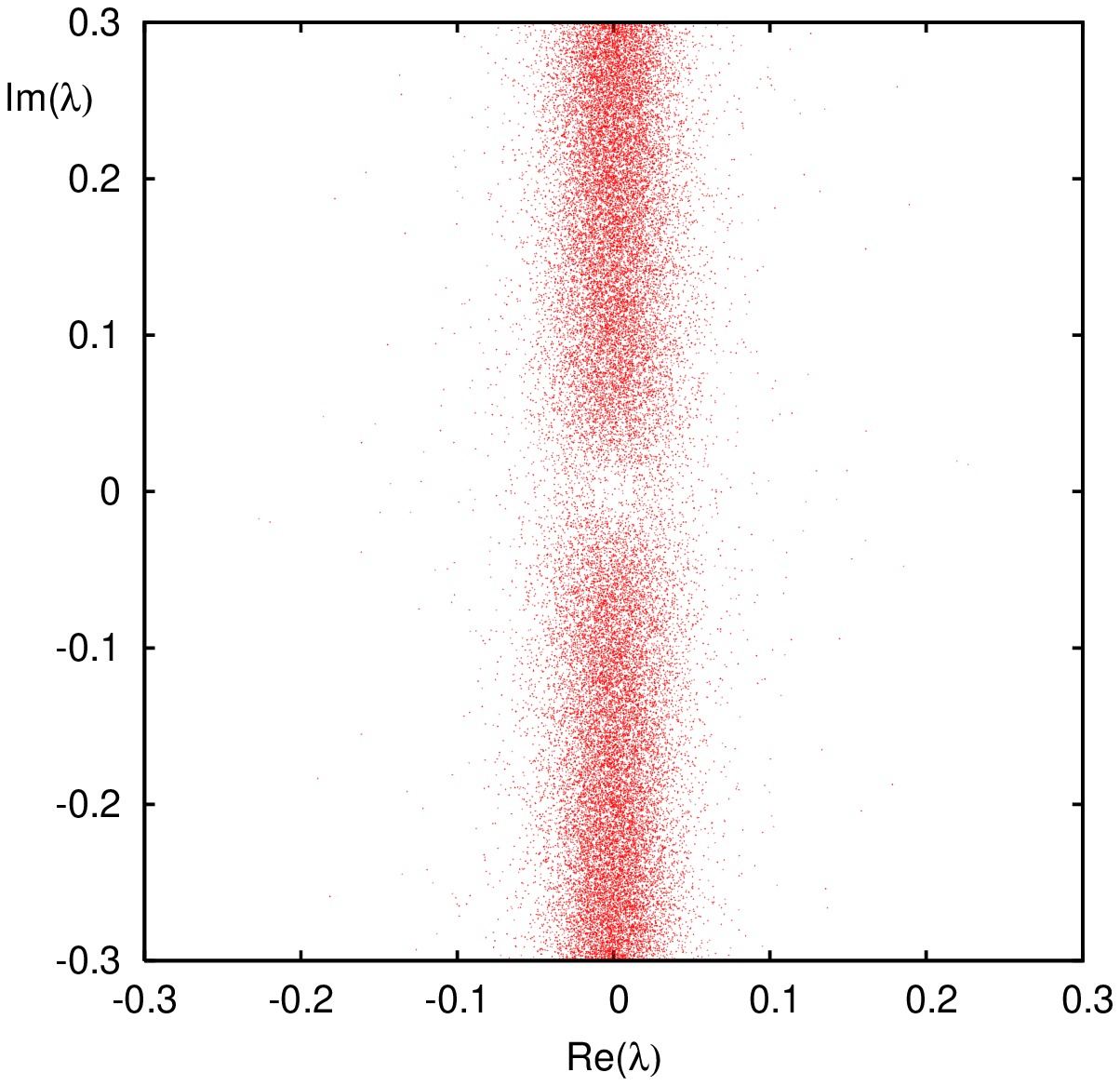}
\hspace{10mm}
\includegraphics[width=5.5cm]{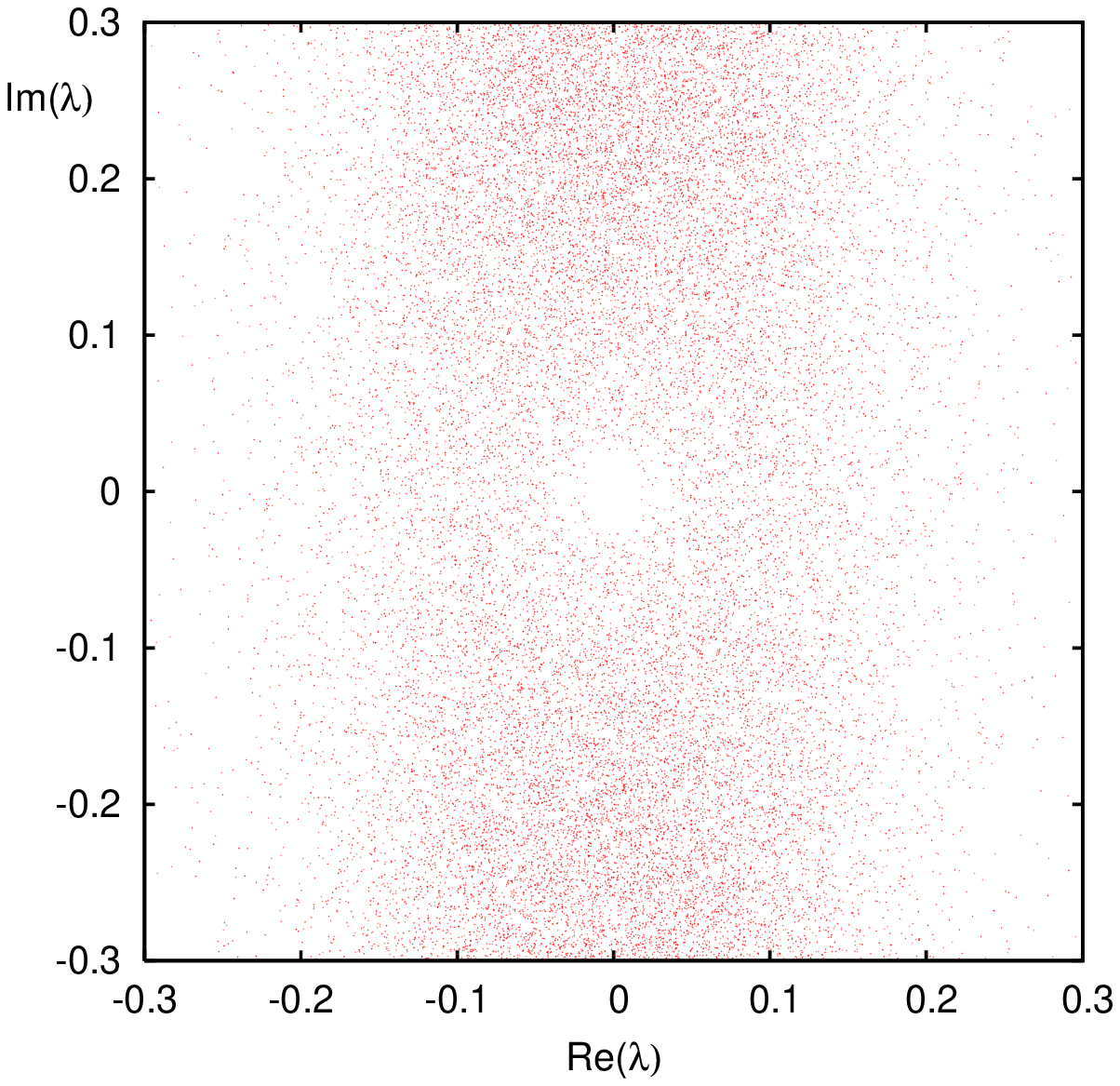}
}
\caption{Scatter plot of the projected eigenvalues $\lambda$ of $D_\text{ov}(\mu)$ for $\mu=0.1$ (left pane) and $\mu=0.3$ (right pane). 
The eigenvalues are projected using $\lambda\to 2\lambda/(2-\lambda)$.
At $\mu=0$ this mapping projects the eigenvalues from the GW-circle to the imaginary axis. 
At $\mu\ne0$, the same mapping projects the eigenvalues onto a band parallel to the imaginary axis.
As $\mu$ is increased the eigenvalues spread further into the complex plane. 
}
\label{Fig:scatter}
\end{figure}

The spectral density of $D_\text{ov}(\mu)$ is given by
$\rho^\text{ov}(\lambda_r,\lambda_i)=\langle\sum_k
\delta(\lambda-\lambda_k)\rangle$ with $\lambda=\lambda_r+i\lambda_i$,
where the average is over configurations.  The distribution of (projected) eigenvalues in the complex plane is shown in Figure~\ref{Fig:scatter}. The claim is that the
distribution of the small eigenvalues of $D_\text{ov}(\mu)$ is
universal and given by RMT.  The chiral RMT model for the Dirac operator is
\cite{Step96}
\begin{align}
  \label{eq:rmt-model}
  D_\text{RMT}(\mu)=
  \begin{pmatrix}0&iW+\mu\\iW^\dagger+\mu&0\end{pmatrix}\:,
\end{align}
where $W$ is a complex matrix of dimension $n\times(n+\nu)$ with no
further symmetries (we take $\nu\ge0$ without loss of generality).
The matrix in Eq.~\eqref{eq:rmt-model} has $\nu$ eigenvalues equal
to zero.  The spectral correlations of $D_\text{RMT}(\mu)$ on the
scale of the mean level spacing were computed in
Refs.~\cite{sv04,o04,aosv05}.  In the quenched approximation, the
result for the microscopic spectral density reads
\begin{align}
  \label{eq:rmt}
  \rho_s^\text{RMT}(x,y) = \frac {x^2+y^2}{2\pi\alpha}
  e^{\frac{y^2-x^2}{4\alpha}} 
  K_\nu\left(\frac {x^2+y^2}{4\alpha}\right)
  \int_0^1 t\, dt\, e^{-2\alpha t^2}|I_\nu(t z)|^2\:,
\end{align}
where $z=x+iy=\lambda\Sigma V$, $I$ and $K$ are modified Bessel functions, and
$\alpha=\mu^2 f_\pi^2 V$.  
$\Sigma$ and $f_\pi$ are low-energy constants that can be obtained
from a two-parameter fit of the lattice data
to the RMT prediction, Eq.~\eqref{eq:rmt}.  (The normalization is fixed
by $\int dx\,dy\,\rho^\text{ov}(x,y)=12V$ and does not introduce
another parameter.)  
For $x\ll\alpha$, Eq.~\eqref{eq:rmt} 
becomes radially symmetric \cite{jac-lectures},
\begin{align}
  \label{eq:strong}
  \rho_s^\text{RMT}(x,y)\to\frac{\xi}{2\pi\alpha}
  K_\nu\left(\xi\right)I_\nu\left(\xi\right)
\end{align}
with $\xi=|z|^2/4\alpha$, and the fit only involves the single parameter $\Sigma/f_\pi$.

\begin{figure*}[-t]
  \hspace*{-2mm}\includegraphics[height=79mm]{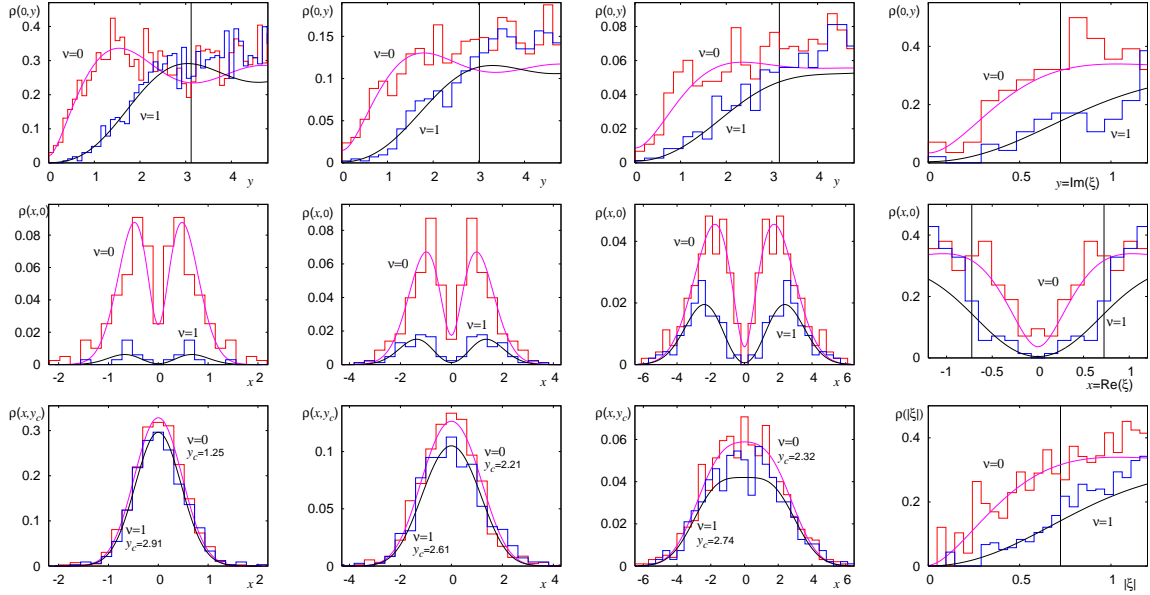}
  \caption{Density of the small eigenvalues  $z=x+iy=\lambda\Sigma V$ of $D_\text{ov}(\mu)$ in the complex plane (after projecting
    $\lambda\to2\lambda/(2-\lambda)$) for
    (from left to right) $\mu=0.1,0.2,0.3,1.0$.  The histograms are
    lattice data for $\nu=0$ and $\nu=1$, and the solid lines are the
    corresponding RMT prediction of Eq.~\protect\eqref{eq:rmt}, integrated
    over the bin size.  Top: cut along the imaginary axis, middle: cut
    along the real axis, bottom: cut parallel to the real axis at $y_c$.  
    Vertical lines indicate the fit interval.
	For $\mu=1.0$ the distribution of the small eigenvalues is radially symmetric up to $|\xi|\sim0.7$, with $\xi=|z|^2/4\alpha$, and therefore only the ratio $\Sigma/f_\pi$ can be determined from a fit to Eq.~\protect\eqref{eq:strong}. In the rightmost bottom plot the data are integrated over the phase.
  }
  \label{fig:rmt1}
\end{figure*}

\begin{table}[t]
\centerline{
\parbox{8.3cm}{
    \begin{tabular}{clclc}
	\hline\hline\\[-4mm]
      $\mu a$ & \multicolumn{1}{c}{$\Sigma a^3$} & $f_\pi a$ &
      \multicolumn{1}{c}{$\Sigma a^3/f_\pi a$} &
      \multicolumn{1}{c}{$\chi^2/\text{dof}$}\\[1mm]\hline \\[-3mm]
      0.0 & 0.0816(6) & \multicolumn{1}{c}{--} &
      \multicolumn{1}{c}{--} & 1.10 \\ 
      0.1 & 0.0812(11) & 0.261(6) & 0.311(5) & 0.67\\
      0.2 & 0.0785(14) & 0.245(5) & 0.320(4) & 0.78\\
      0.3 & 0.0824(17) & 0.248(5) & 0.332(4) & 1.03\\
      1.0 & \multicolumn{1}{c}{--} & \multicolumn{1}{c}{--} &
      0.603(18) & 0.42 \\
	  \hline\hline
    \end{tabular}
  \caption{Fit results for $\Sigma$ and $f_\pi$.
For $\mu=1.0$ only $\Sigma/f_\pi$
can be determined (see Figure~\protect\ref{fig:rmt1}).  
  \vspace{-4mm}}
  \label{table2}
  }}
\end{table}

In Fig.~\ref{fig:rmt1} we compare our lattice data to the RMT
prediction.  We display various cuts of the eigenvalue density in the
complex plane as explained in the figure captions.  The data agree
with the RMT predictions within our statistics.  $\Sigma$ and $f_\pi$
were obtained by a combined fit to the $\nu=0$ and $\nu=1$ data for
all three cuts and are displayed in Table~\ref{table2}.  (These
numbers have no physical significance at $\beta=5.1$.)  

In summary, we have shown how to include a quark chemical potential in
the overlap operator.  The operator still satisfies a Ginsparg-Wilson
relation and has exact zero modes.  The distribution of its small
eigenvalues agrees with predictions of non-Hermitian RMT for trivial
and nontrivial topology.  Our initial lattice study should be extended
to weaker coupling, larger lattices, and better statistics.  Work on
approximation methods to enable such studies is in progress.  For
small volumes, reweighting with the fermion determinant should allow
us to test RMT predictions for the unquenched theory \cite{AB06}.

This work was supported in part by DFG.  
The simulations were performed on a \mbox{QCDOC} machine in Regensburg 
using USQCD software and Chroma \cite{usqcd}, and GotoBLAS optimized for QCDOC.

\end{document}